\documentclass[runningheads]{llncs}
\usepackage{graphicx}
\usepackage{amsmath}
\usepackage{amssymb}
\begin{document}
\title{METGAN: Generative Tumour Inpainting and Modality Synthesis in Light Sheet Microscopy}%
\titlerunning{METGAN}
\author{Izabela Horvath \inst{1,2} \and       
Johannes Paetzold \inst{1,2,3} \and 
Oliver Schoppe \inst{1,2} \and 
Rami Al-Maskari \inst{1,2} \and 
Ivan Ezhov\inst{1,3} \and 
Suprosanna Shit \inst{1,3} \and 
Hongwei Li \inst{1,4} \and 
Ali Ertuerk \inst{2} \and 
Bjoern H. Menze \inst{1,4}
} 

\authorrunning{Horvath and Paetzold et al.}

\institute{Department of Computer Science, Technische Universität München, Germany
\and Institute for Tissue Engineering and Regenerative Medicine (ITERM), Helmholtz Zentrum München, Neuherberg, Germany
\and TranslaTUM Center for Translational Cancer Research, München, Germany
\and Department of Quantitative Biomedicine, University of Zürich, Zurich, Switzerland \\
\email{izabela.horvath@tum.de johannes.paetzold@tum.de bjoern.menze@tum.de}
}

\maketitle              % typeset the header of the contribution
\begin{abstract}
Novel multimodal imaging methods are capable of generating extensive, super high resolution datasets for preclinical research. Yet, a massive lack of annotations prevents the broad use of deep learning to analyze such data. 
So far, existing generative models fail to mitigate this problem because of frequent labeling errors. 
In this paper, we introduce a novel generative method which leverages real anatomical information to generate realistic image-label pairs of tumours. We construct a dual-pathway generator, for the anatomical image and label, trained in a cycle-consistent setup, constrained by an independent, pretrained segmentor. The generated images yield significant quantitative improvement compared to existing methods. To validate the quality of synthesis, we train segmentation networks on a dataset augmented with the synthetic data, substantially improving the segmentation over baseline.

\end{abstract}
%
%
%
%auto-ignore
\section{Introduction}

Recently, the combination of fluorescence microscopy and tissue clearing has enabled the generation of single-cell resolution, terabyte sized 3D datasets of whole specimens and organs \cite{susaki2014whole,ueda2020tissue,zhao2020cellular}. These datasets facilitate the quantitative study of disease or age-induced anatomical alterations, as well as drug targeting, in human organs or animal models. An intriguing feature of such datasets is the multi-channel nature of the data. An autofluorescence channel, denoted as "anatomical channel", images general tissue, and a pathology channel marks objects of interest such as tumours ("tumour channel") using fluorescence dyes. The sheer size and multi-channel characteristics of these data require the use of high throughput deep learning methods to analyze and segment them \cite{todorov2019automated,pan2019deep,shit2020cldice}. However, data volume and complexity increase the manual annotation cost, as the timeframe required for manual labeling of a single scan of a whole mice can span up to two months. This evidently motivates the development of new approaches for synthetic image-label pair generation and data augmentation \cite{paetzold2019transfer}. Nevertheless, existing methods often fail to generate semantically correct images when underlying annotations used for training are noisy. Moreover, current applications that focus on the semantics fail at taking advantage of existing prior anatomical information, generating semantically correct, but often trivial images.
\begin{figure}
\includegraphics[width=0.95\textwidth]{ 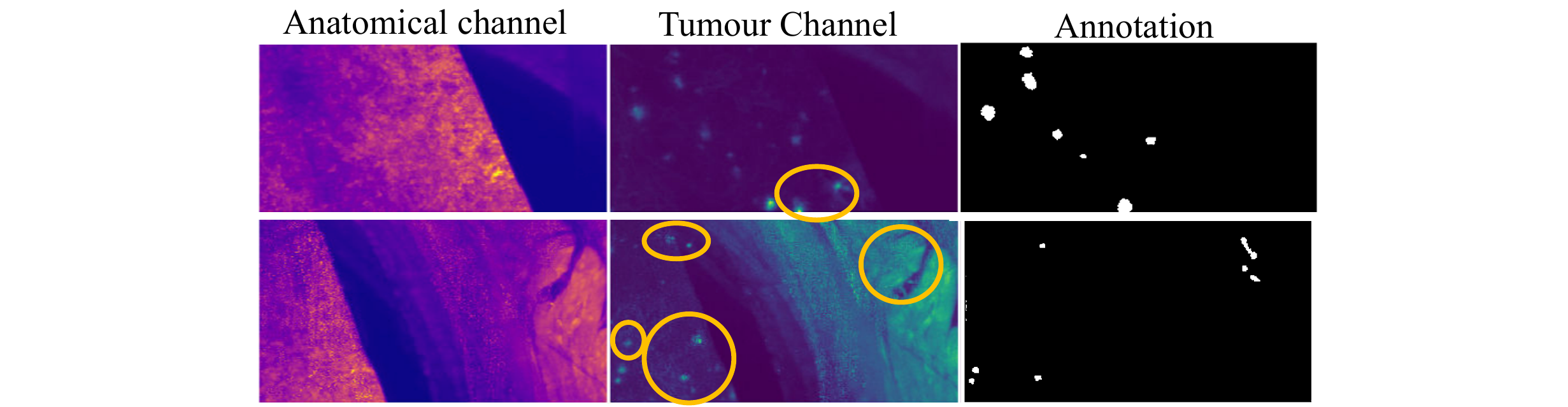}
\caption{%\small 
Motivation: the characteristic dim metastases can only be identified by multiple adjustments of contrast, and only from certain angles. Such properties lead to inconsistent annotations, even from experts. First, these inconsistencies make the labeling very expensive. Second, training segmentation networks on these datasets is only stable for large numbers of annotated samples. Both of these aspects motivate the need for our generative tumour inpainting to create labeled datasets.}
\label{fig_motivation}
\end{figure}

This paper presents a novel generative adversarial network (GAN) based approach, which leverages the advantages of fluorescence microscopic datasets, namely a high SNR  and multiple channels, to generate realistic data. 
Our application focuses on a dataset pertaining metastatic spread in a whole mouse organism. Based on information in the anatomical channel, we generate synthetic images in the pathology (cancer) domain, with objects of interest (tumours) placed in user-defined locations. 
Thus, we derive the generated samples from existent and distinctive priors, without the additional burden on the generator of having to synthesise both diverse backgrounds and foreground.
Additionally, our method solves  characteristic inconsistencies in foreground data and labels (see Figure \ref{fig_motivation}).\\

 The use of generative adversarial networks for synthetic data generation and augmentation \cite{shorten2019survey} is a current topic of interest for improvements in segmentation and classification tasks. Goodfellow \emph{et al.} introduced the concept of GANs \cite{goodfellow2014generative}, which was successfully improved and extended towards image-to-image translation \cite{isola2017image,zhu2017unpaired,choi2018stargan}, among other applications in medical imaging \cite{yi2019generative,kazeminia2020gans,9301322}.
For medical images, label-based conditional GANs have been used for data infill \cite{li2019diamondgan} and augmentation for underrepresented classes \cite{mariani2018bagan}, to improve segmentation, \cite{zhang2018translating} as well as classification \cite{vandenhende2019three,chaitanya2019semi,schlegl2017unsupervised}. 
Semantic image synthesis has been enhanced with  spatially-adaptive normalization \cite{park2019semantic}. Related to our oncology application in lesion segmentation, Qasim\emph{et al.} introduced a third player in the generative game, which improved performance of the downstream segmentation task \cite{qasim2020red}. 
Cohen \emph{et al.} showed that using distribution matching losses can lead to hallucinating tumours, which translated to medical misdiagnosis \cite{cohen2018distribution}, clearly showing the need for advanced image synthesis techniques, trained with losses that punish the generation of unwanted structures. \\
 
\noindent \textbf{\textit{Contributions: }} 1) We develop a novel generative model, MetGAN, to synthesize realistic microscopic images of tumour metastases, based on real autofluorescence image information and arbitrarily placed tumour labels. Our model consists of a dual-pathway generator, MetGen, trained in a cycle-consistent setup, and further constrained by an independent, pretrained segmentor. Our novelty lies in: the generator architecture, the addition of a passive segmentor in the cycle-consistent training setup, and the further constraint with a pair-wise loss for improved domain adaptation.
2) We present qualitative results of our generated microscopic tumour images, and we quantitatively evaluate the error comparative to the real ground truth images. This shows the superiority of our model over existing state of the art methods.  
3) We extensively validate our method in a downstream segmentation task, where we use our generated tumour images to train a segmentation network. We show that training solely on generated image-label pairs achieves identical performance as training on a large set of real data. Furthermore, augmentation of the real dataset with synthtetic data improves lesion detection.

%auto-ignore
\section{Methods}
\subsubsection{Architecture}
\begin{figure}[h!]
\includegraphics[width=\textwidth]{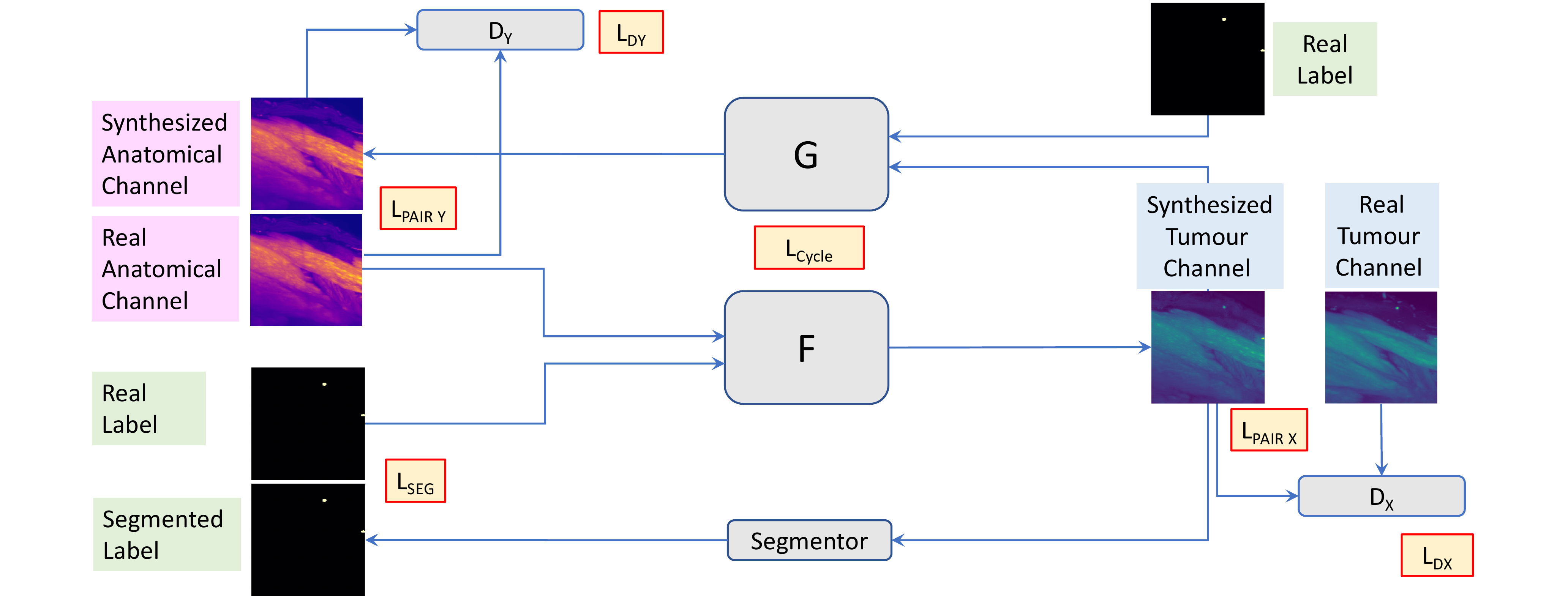}
\caption{%\small 
Proposed conditional GAN training setup: %with a segmentation network. 
The generator F %$G_{X \to Y}$ 
learns the mapping from the anatomical to the tumour channel, conditioned on the imposed label, through the discriminator $D_X$. Because we train in a cycle consistent manner, G
%$G_{Y \to X}$ 
learns the inverse mapping, through %the discriminator 
$D_Y$. A fixed pretrained segmentor is used to enforce the desired semantics %labels 
by punishing  F %$G_{X \to Y}$ 
for incorrect tumour placement. } \label{fig_arch}
\end{figure}

In this work, we propose a cycle-consistent 2D framework for domain adaptation and semantic tumour inpainting using a customized GAN, whose architecture is depicted in Figure~\ref{fig_arch}. Our setup is inspired by CycleGAN, a proven model in domain translation tasks \cite{zhu2017unpaired}.\\

It has been shown that networks which employ simple batch normalization layers tend to lose semantic information when it comes to label-based generation  \cite{park2019semantic}. This aspect motivated us to construct proposed generator network, \textit{MetGen}, with an additional pathway, tailored for semantic synthesis. Thus, we delimit two paths: one following a traditional U-net architecture \cite{Vugt2019}, that receives the anatomy channel as input; and a second path composed of 7 Spade ResNet Blocks\cite{park2019semantic}, which process the label-input to the network.
We then merge the resulting features of the two paths in the upconvolutional layers of the U-net decoder (see Supplementary). This separation better preserves the flow of semantic information and results in a more accurate label-based inpainting compared to a naive channel-wise concatenation of the input image and the annotation.\\

Furthermore, we want to ensure that the generated output is consistent with the desired label - a constraint we enforce through a pre-trained and frozen segmentor network, employed as a passive player in the training process. Its role is twofold: not only does it enforce the placement of metastases at the desired locations, but it also helps to suppress hyperintensities in the anatomy channel which are preserved in the case of CycleGAN, creating semantic ambiguities.

\noindent\textbf{Training Losses}
As presented in Figure \ref{fig_arch}, our final loss consists of four terms. Firstly, we define the discriminator and cycle consistency losses similar to  \cite{zhu2017unpaired}: 

\begin{equation*}
\begin{aligned}
   L_{D}(&F,G,D_X,D_Y,X,Y,L) = \mathbb{E}_{y\sim p_{(y)}}[logD_{Y}(y)] + {E}_{x,l\sim p_{(x,l)}}[log(1-\\&D_{Y}(F(x,l)] 
   + {E}_{x\sim p_{(x)}}[logD_{X}(x)] + {E}_{y,l\sim p_{(y,l)}}[log(1-D_{X}(G(y,l)]
\end{aligned}
\end{equation*}
\begin{equation}
   L_{Cycle}(F,G) = 
   \mathbb{E}_{x\sim p_{(x)}} [\|G(F(x,l),l)-x\|_1] + \mathbb{E}_{y\sim p_{(y)}}[\|F(G(y,l),l)-y\|_1],
\end{equation}

where we denote: X - the anatomical domain, Y - the tumour domain, L - the binary domain marking the presence or absence of a tumour; {\it x,l} ${\sim p_{(x,l)}}$ and  {\it y,l}${\sim p_{(y,l)}}$ are samples from domain X and Y; ${F(x,l)}$ is a mapping from X $\times$ L $\to$ Y; and ${G(y,l)}$ a mapping from Y $\times$ L $\to$ X, with D$_{X}$ and D$_Y$ as corresponding discriminator functions. %s.

Secondly, in order to ensure the desired segmentation map is respected, and that the setup is robust against artefacts and hyperintesities in the autofluorescence channel, we use 
a weighted binary cross-entropy loss $L_{Segm}(F,L)$, given by the predictions of the segmentor network, penalizing the discrepancy between the real and segmented label. % $L_{Segm}$ is .
Lastly, to leverage the paired nature of our data and to facilitate feature translation achieved in the domain adaptation task, we also use a pair-wise loss between real and generated images in each domain:
\begin{equation}
   L_{Pair}(F,G)  = \mathbb{E}_{x,y}[\|F(x,l) - y\|_1] +  {E}_{x,y}[\|G(y,l) - x\|_1].
\end{equation}
The final loss is a weighted linear combination of these components, where the parameters $\alpha_{1-4} $ are hyperparameters adjustable per dataset.
\begin{equation}
   L_{final}(F,G,X,Y,L) = \alpha_1 L_{D}  + \alpha_2 L_{Cycle}  + \alpha_3 L_{Segm}+ \alpha_4 L_{Pair}.
\end{equation}

\section{Experiments}
\noindent\textbf{Dataset}:
In our evaluation, we use a publicly available light sheet microscopy dataset \cite{pan2019deep,schoppe2020deep}. The dataset contains 3528, 300x300 pixel images, with their corresponding ground truth annotations. We selected 1602 samples depicting areas within the bodies and resized them to 256x256 pixels. We put aside 20 \% of the data as a test set that is unseen by any network. The test set includes cases with and without metastases, with a balanced organ-based distribution. Normalization to [-1,1] and random rotations are used at train time.

We compare our proposed solution to established baseline methods such as Pix2Pix and CycleGAN, as well as  mixture models: Pix2Pix and CycleGAN trained with an additional segmentation loss (referred to as "Pix2PixSeg" and "CycleGanSeg"). 
As our generator is agnostic to the discriminator achitecture and training method, we set out to train  MetGen  with a conditional discriminator, as in \cite{isola2017image}, and with a cycle-consistent setup, as in \cite{zhu2017unpaired} (denoted "MetGenCondSeg+" and "MetGAN-", respectively). The difference between "MetGAN-" and "MetGAN" stands in the use of the pair-wise loss $L_{Pair}$. 
We compare the generated images from a qualitative and a quantitative point of view, based on the unseen test set. For this, we use 414 real pairs of anatomical channel, real  label, and tumour channel. Lastly, we implement a downstream study, where we train segmentation networks on synthetic, real and mixed data. \\

\noindent\textbf{Implementation details}: 
We train our models using Pytorch, on an NVIDIA GeForce RTX 2080, for 200 epochs, with batch size=1, using Adam optimizer with $\beta_{1}$ = 0.5, $\beta_{2}$ = 0.999, initial learning rate = 0.0002 and linear learning rate decay after 100 epochs. The pre-trained segmentor network (U-net) developed by Pan \textit{et al.} in \cite{pan2019deep} is loaded, and not modified during training. As a discriminator, we use PatchGAN with 3 layers and a field of view of 70x70 pixels, in a simple or conditional setup (with its vote based on the generator's output for the cycle consistent setup, or concatenation of output and input, otherwise). Training our final method takes approximately 24 hours.
For the final loss function, we use $\alpha_1 = 1$, $\alpha_2 = 10$, $\alpha_3 = 100$, and $\alpha_4 = 10$, obtained empirically. The baseline methods were trained using default parameters proposed by \cite{zhu2017unpaired}.
%auto-ignore
\section{Results}
\begin{figure}[ht!]

\centering
\includegraphics[width=0.85\textwidth]{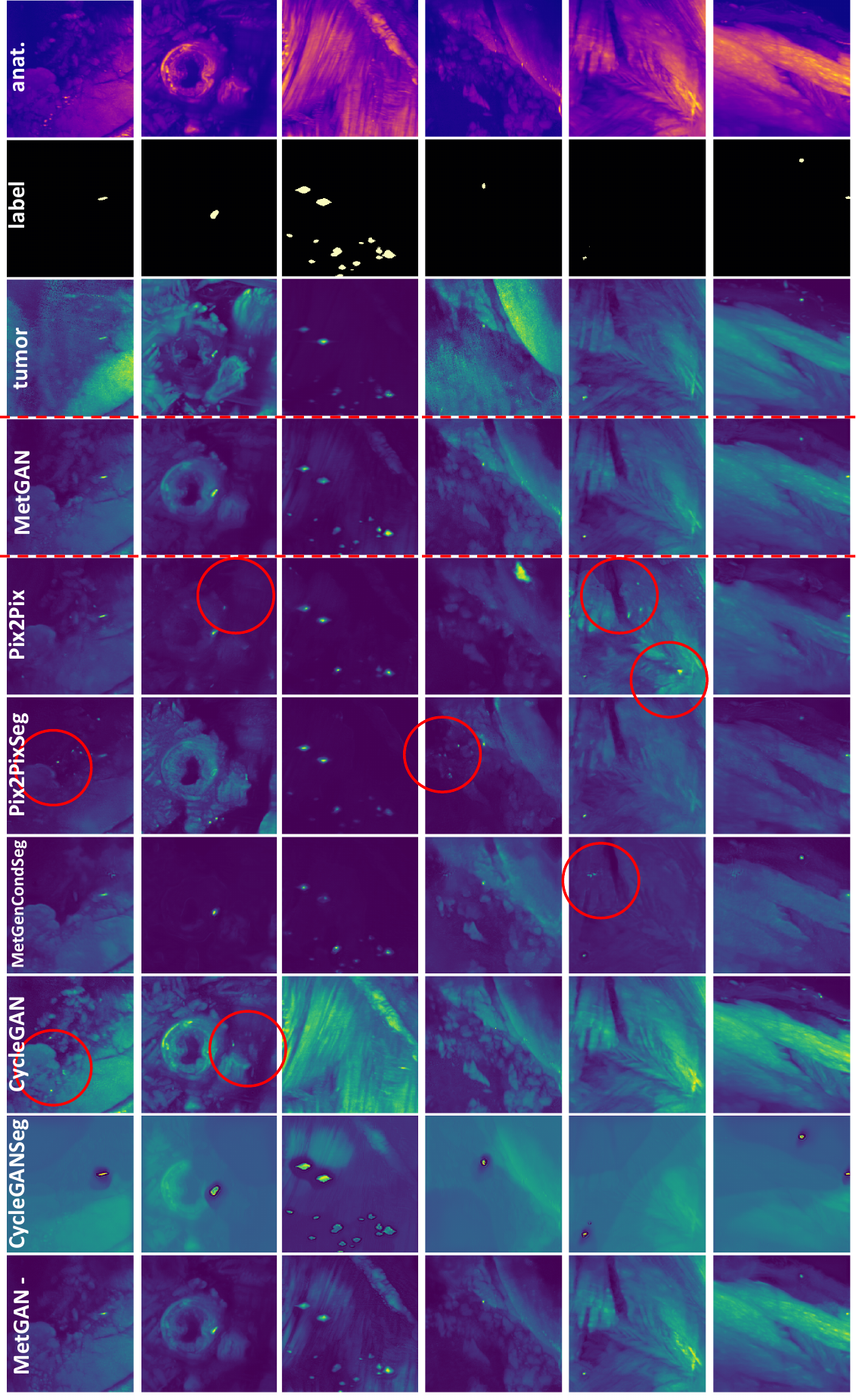}
\caption{\scriptsize{Qualitative results from our proposed generative method compared to multiple state of the art methods: Pix2Pix, Pix2PixSeg, MetGenCondSeg, CycleGAN, CycleGANSeg, MetGAN-. We find that our method generates the most realistic looking tumour images, with tumours in the correct locations. The state of the art fails at respecting the imposed semantic map and lacks in the domain adaptation, either by not using features from the anatomical domain or by keeping too many.}} \label{fig4}
\end{figure}

\textbf{Quantitative Results}:
In Table \ref{table_quant}, we compare the generated "tumour channel" images to the corresponding ground truth using three image similarity metrics. % We observe that MetGAN achieves the best scores across all tested architectures. 
Our first implementation, %the conditional generation with an added segmentor (Pix2PixSeg),
Pix2PixSeg, learns to produce structures at the desired locations (see Figure \ref{fig4}), improving image similarity compared to the basic Pix2Pix and CycleGAN. However, using MetGen as a here generator decreases the performance in regards to MAE, MSD and SSIM. In the cycle consistent setup, CycleGAN produces good similarity scores, but fails to respect the semantic map. Here, adding the segmentor (CycleGANSeg) weakens the performance substantially. MetGAN- improves MAE and MSD over CycleGAN; however, the SSIM is worse than in CycleGAN. 
Finally, our method leads to a consistent improvement of the generated images compared to the state of the art according to the three metrics.

\begin{table}[ht]
\centering
\caption{%\small
In a quantitative comparison, our proposed method outperforms state of the art methods in mean absolute error (MAE), mean sum of squared differences (MSD) and structural similarity index measure (SSIM).Our improvements are all significant based on t-test analysis (all p-values ${<}$0.005). Best scores are indicated in bold digits.}\label{table_quant}
\begin{tabular}{|l|l|l|l|}
\hline
Network &  MAE$\downarrow$ & MSD$\downarrow$ & SSIM $\uparrow$ \\
\hline
MetGAN & \textbf{0.1111}& \textbf{0.0231} & \textbf{0.7003}\\
Pix2Pix &  0.1221 & 0.0281 & 0.6506 \\
Pix2PixSeg &  0.1184 & 0.0260 & 0.6585\\
MetGenCondSeg & 0.1257 & 0.0286 & 0.6434\\
CycleGAN & 0.1296 & 0.0312 & 0.6560 \\
CycleGANSeg & 0.2100 & 0.0591 & 0.5633 \\
MetGAN -  & 0.1203 & 0.0265 & 0.6540  \\
\hline
\end{tabular}
\end{table}

\noindent\textbf{Qualitative Results}:
Figure \ref{fig4} shows examples of qualitative results obtained for our method and baselines. We observe the following common features: while standalone Pix2Pix and CycleGAN obtain good image similarity and domain adaptation, they either fail to create the desired tumours, or the networks hallucinate features in undesired places. Adding a segmentor network helps to ensure that the new structures are inpainted at the desired location, but still leads to additional unwanted tumours (Pix2PixSeg), or fails on the domain adaptation task (CycleGANSeg). The use of our generator improves the synthesis in both conditional (MetGenCondSeg) and cycle-consistent setups (MetGAN-). Nevertheless,  MetGenCondSeg often fails to maintain features from the anatomical channel, producing dark images with bright metastases, which is a simple, but only occasionally correct solution to the task. %However, most often it does not capture the whole distribution of the data. 
On the other hand, MetGAN- keeps too many of the original features. An optimal balance is reached with our proposed solution, MetGAN, which consistently places tumor at the imposed label location, whilst adapting features from the anatomical domain.\\

\begin{figure}[ht!]
    \centering

    %\rule{1cm}{.6cm}
   \includegraphics[width=0.7\textwidth]{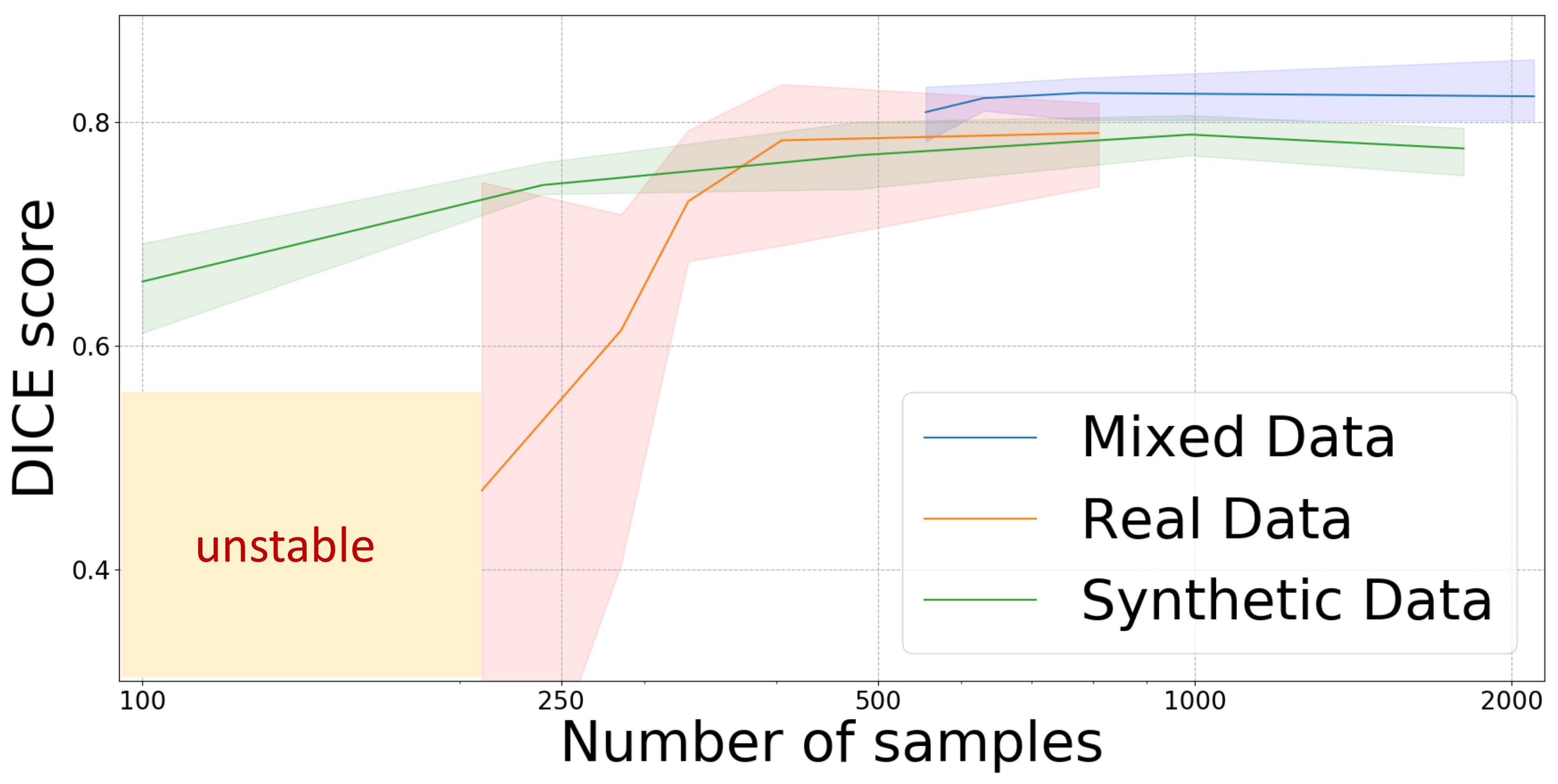}
    
\caption{Segmentation performance as a function of dataset size during training. Mean values as lines (see Table), and minimum and maximum achieved during  5-fold cross validation as delimiters of the areas. We can observe that the inclusion of synthetic data makes training the segmentation network more stable; especially in the case of small datasets, where training on real data is unstable and volatile. On the other hand the performance obtained with purely synthetic or mixed data is more stable.}\label{fig_perf_DICE}
\end{figure}

\noindent\textbf{Downstream segmentation task analysis}:
We analyze how \textit{MetGAN} images can be used to augment real data for training segmentation networks. We train segmentation models on sets of synthetic, real or mixed data, with a varying number of samples from each set. For generating synthetic data, we use randomized unpaired combinations of real anatomical channel images, and real non-zero labels or combinations thereof (e.g. 2 non-zero labels merged together, in order to create more objects of interest in one image). As a segmentor network, we use the U-net developed in \cite{pan2019deep}. We perform 5-fold cross-validation for all experiments. We evaluate the segmentation performance using lesion-wise Dice, precision, recall and Jaccard Index, on an unseen test dataset (which is not used in GAN training or image generation either), see Figure \ref{fig_perf_DICE}. \\

\noindent \textbf{Purely synthetic data}: By training a segmentation network only with synthetically generated data, we observe that we achieve a similar performance (78.8\% Dice) as by using real data (79.8\% Dice). Moreover, for a low number of samples, we not only outperform the model trained with real data, but also offer better training stability with decrease in dataset size. We attribute this to the fact that our approach can generate images with an increased number of objects of interest, allowing us to have a more representative depiction of the data distribution, even in a low sample regime.  Nevertheless, increasing the amount of training images past a certain point ($\approx$1000 samples, in our case) results in a slight decrease of performance, as the network overfits on synthetic data.\\

\noindent \textbf{Mixed data}: Compared to training the segmentor purely on real images, we observe that our augmentation can increase the performance. Our experiments highlight that, by using as little as 25\% of the available real data, together with synthetically generated samples, we can outperform the real data baseline. Adding our synthetic data improves the performance up to a plateau point, where we speculate that the limitations are caused by the inherent variability in annotation quality, see Figure \ref{fig_motivation}. Past this point, adding more generated samples leads to overfitting and a slight decrease in performance. On a per tumour basis, our data augmentation increases the mean number of detected metastases (true positives) from 83 to 95, whilst simultaneously decreasing the number of false positives from 41 to 21. The detection is especially improved for small-sized or dim tumours located in the lungs, showing that our network can produce diverse objects that can be used to improve difficult cases. \\

\begin{table}[ht]
\centering
\caption{Mean segmentation performance of models trained on real (R), synthetic (S), and mixed data. We see that training based on synthetic data reaches a performance similar to the real data, even outperforming real data at low numbers of training samples. The best segmentation scores can be obtained by combining real and synthetic data.} \label{table_ablation}    
\begin{tabular}{|l|c|c|c|c|}
    \hline
    
    Samples                                                               & DICE$\uparrow$           & Prec.$\uparrow$                & Rec.$\uparrow$                 & J. I.$\uparrow$              \\ \hline
    100 synthetic                                                         & 0.657                           & 0.607                           & 0.718                           & 0.489                           \\
    120 synthetic                                                         & 0.680                           & 0.611                           & 0.770                           & 0.515                           \\
    240 synthetic                                                         & 0.744                           & 0.711                           & 0.783                           & 0.592                           \\
    480 synthetic                                                         & 0.770                           & 0.767                           & 0.776                           & 0.626                           \\
    1000 synthetic                                                        & 0.788                           & 0.776                           & 0.804                           & 0.650                           \\
    1800 synthetic                                                        & 0.776                           & 0.780                           & 0.773                           & 0.640                           \\ \hline
    200 real(25\%)                                                       & 0.470                           & 0.465                           & 0.538                           & 0.307                           \\
    280 real(35\%)                                                       & 0.614                           & 0.684                           & 0.652                           & 0.443                           \\
    400 real(50\%)                                                       & 0.784                           & 0.810                           & 0.764                           & 0.645                           \\
    810 real(100\%)                                                      & 0.790                           & 0.800                           & 0.781                           & 0.653                           \\ \hline
    200R+500S    & 0.809                           & 0.809                           & 0.811                           & 0.679                           \\
    400R+500S    & 0.819                           & 0.812                           & \textbf{0.828} & 0.693                           \\
    810R+500S    & \textbf{0.826} & \textbf{0.842} & 0.811                           & \textbf{0.704} \\
    810R+1500S   & 0.823                           & 0.823                           & 0.824                           & 0.699                           \\ \hline
\end{tabular}
\end{table}

\clearpage
\section{Conclusions}
 In this paper, we introduce a novel generative method, which is able to leverage real anatomical information to generate realistic image-label pairs of tumours. We achieve this by implementing a dual-pathway generator, for the anatomical image and label, trained in a cycle-consistent fashion, which is constrained by an independent, pretrained segmentor. We generate images which are substantially more realistic in terms of quantitative and qualitative results, compared to 5 different state of the art models. Moreover, we train segmentation networks on real, generated and mixed data. We find that data synthesized with our method improves segmentation; both from a training stability point of view, observable at low data regimes; as well as from a lesion-detection point of view. Using our method leads to higher segmentation scores when used to augment real data, and can potentially be further exploited by focusing on underrepresented or low-performance cases.
 
\pagebreak

{\bibliographystyle{splncs04}
\bibliography{main}
}
\pagebreak

%auto-ignore
\section{Supplementary Material}

\begin{figure}[ht]
\includegraphics[width=\textwidth]{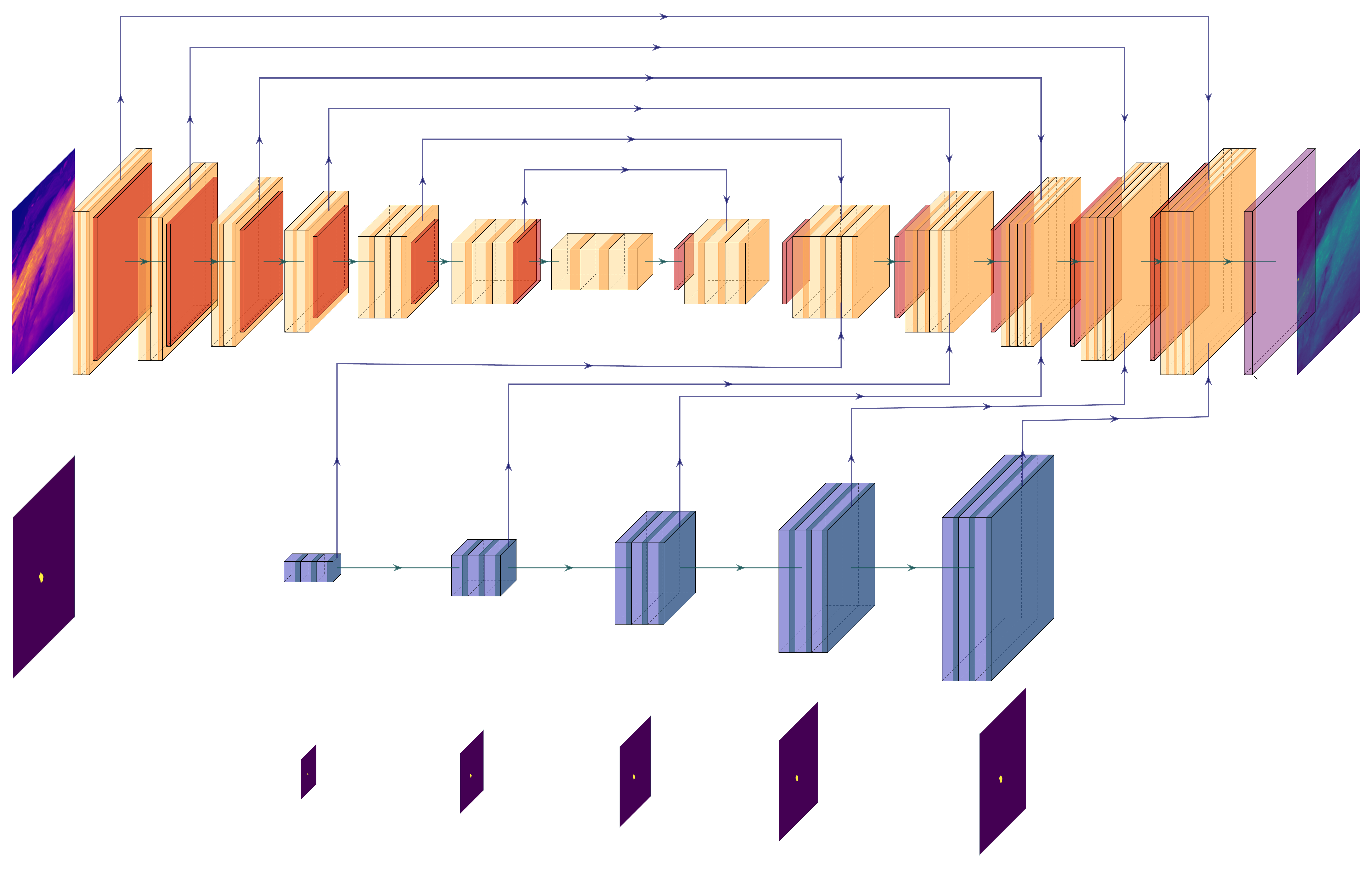}
\caption{Depiction of the proposed generator architecture, \textit{METGEN}. Our setup processes the input image through a U-net architecture, and the semantic information through Spade ResNet Blocks.The features are concatenated final upconvolutional layers and used for spatially and semantically accurate inpainting.} \label{fig_gen}
\end{figure}

\begin{figure}[ht]
\includegraphics[width=\textwidth]{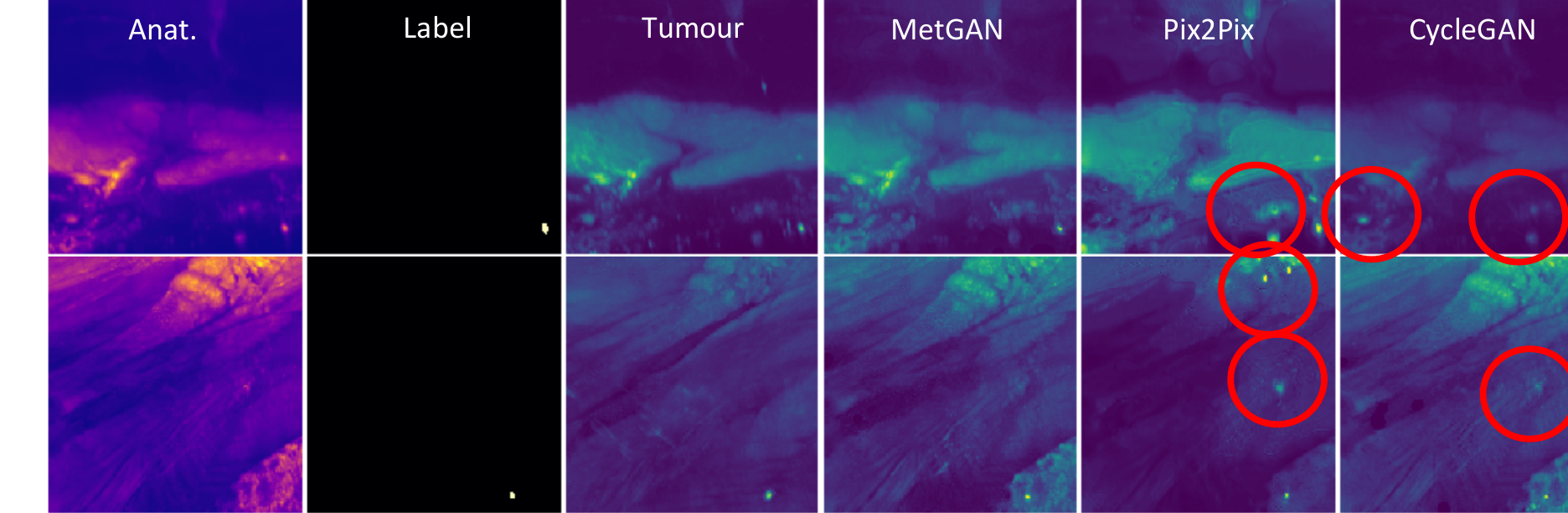}
\caption{Robustness testing of our method. We train \textit{MetGAN}, Pix2Pix and CycleGAN with data containing 25\% shuffled labels. We can observe that MetGAN is robust to label inconsistency, and produces realistic images that are in line with the input to the network.} \label{fig_gen1}
\end{figure}

\begin{figure}[ht]
\includegraphics[width=1\textwidth]{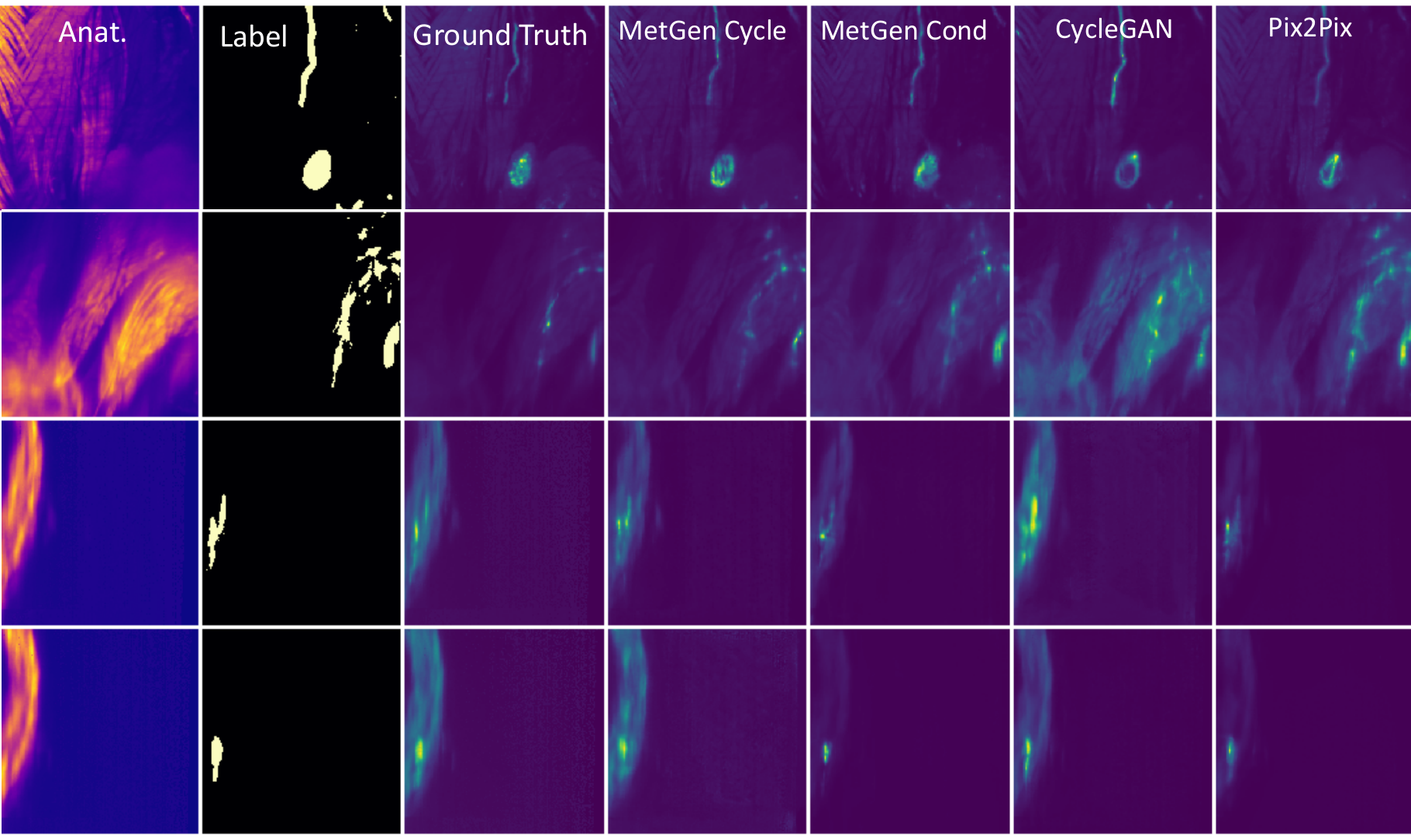}
\caption{Validation of our generator, \textit{MetGEN} (no segmentor) on a second proprietary fluorescence microscopy dataset of the mouse peripheral nervous system. The anatomy channel is translated to contrast-enhanced nerve channel, with semantics based on the imposed annotation. We trained MetGen in a Cycle-consistent setup (MetGen-Cycle) or with a conditional discriminator (MetGen Cond). We compare with CycleGAN and Pix2Pix. 10500 images were used for training, and the networks were tested on 2000 images. We can observe that the best similarity between generated and ground truth image is obtained by our setup.} 
\end{figure}

\begin{table}[ht]
\centering
\caption{Qunatitative validation of our generator, MetGen, (without a segmentor) versus the state of the art, on a second, proprietary dataset pertaining a mouse peripheral nervous system (as described above). Mean absolute error (MAE), mean sum of squared differences (MSD), structural similarity index measure (SSIM), and cross-correlation coefficient (CCoeff) are used for comparing generated images vs ground truth. Best scores are indicated in bold digits and are statistically significant with a t-test based p-value $<$ 0.005.}\label{table_quant_nerve}
\begin{tabular}{|l|l|l|l|l|}
\hline
Network &  MAE$\downarrow$ & MSD$\downarrow$ & SSIM $\uparrow$& CCoeff$\uparrow$\\
\hline
MetGen Cycle & \textbf{0.075}& \textbf{0.016} & \textbf{0.656} & \textbf{0.872}\\
MetGen Cond &  0.078 & 0.018 & 0.643 & 0.848\\
Pix2Pix &  0.081 & 0.019 & 0.629 & 0.85\\
CycleGAN & 0.091 & 0.024 & 0.614 & 0.851\\

\hline
\end{tabular}
\end{table}

\end{document}